\documentclass[a4paper,11pt]{article}
\usepackage{theorem}
\theorembodyfont{\itshape}
\newtheorem{pro}{Proposition}

\usepackage{amssymb}
\usepackage{latexsym}
\usepackage{amsmath}
\usepackage{braket}
\usepackage[dvips]{graphicx}
\usepackage{cite}
\title{Uniqueness of Nash equilibria in quantum Cournot duopoly game}
\author{Yohei Sekiguchi,$^1$ Kiri Sakahara,$^1$ Takashi Sato $^{1,2}$\\
$^1$ {\normalsize\textit{Graduate School of Economics, University of Tokyo, 7-3-1 Hongo, Bunkyo-ku, Tokyo, Japan}}\\
$^2$ {\normalsize\textit{Faculty of Economics, Toyo University, 5-28-20 Hakusan, Bunkyo-ku, Tokyo, Japan}}
}

\begin{document}
\maketitle

\begin{abstract}
A quantum Cournot game of which classical form game has multiple Nash equilibria is examined.
Although the classical equilibria fail to be Pareto optimal, the quantum equilibrium exhibits the following two properties, (i) if the measurement of entanglement between strategic variables chosen by the competing firms is sufficiently large, the multiplicity of equilibria vanishes, and, (ii) the more strongly the strategic variables are entangled, the more closely the unique equilibrium approaches to the optimal one.

PACS numbers: 03.67.-a, 02.50.Le
\end{abstract}

\section{Introduction}
 Game theory is a powerful mathematical tool to analyze various natural and social phenomena \cite{Gibbons, Axelrod, Maynard_Smith}.
 After the publication of Meyer \cite{M99}, there has been a great deal of effort to extend the classical game theory into the quantum domain, and it has been shown that quantum games may have significant advantages over their classical counterparts \cite{M99, EWL99, BH01}.
 The classical game theory includes a fatal drawback, namely, the multiplicity of equilibria.
 Battle of sexes, chicken game, and stag hunt are famous examples of games with multiple equilibria.
 For a game possessing multiple equilibria, the classical game theory can say nothing about the predictability of the outcome of the game: there is no particular reason to single one out of these equilibria.
 Until the present, several quantum extensions are considered to resolve this problem, e.g., battle of sexes \cite{A05, B00, Du01, Du00, MW00a, MW00b, NT04}, chicken game \cite{EW00, FH07, IT07, ITC08}, and stag hunt \cite{IT07, ITC08, T03}.
 We also attack this problem by analyzing a quantum extension of a game which describes market competition.
 
 In economics, many important markets are neither perfectly competitive nor perfectly monopolistic, that is, the action of individual firms affect the market price \cite{Tirole}.
 These markets are usually called \textit{oligopolistic} and can be analyzed based on Game theory.
 Recently, Li et al. \cite{LDM02} investigated the quantization of games with continuous strategic space, a classic instance of which is the Cournot duopoly \cite{Cournot}, in which firms compete on the amount of output they will produce, which they decide on independently of each other and at the same time.
 Li et al. \cite{LDM02} showed that the firms can escape the frustrating dilemma--like situation if the structure involves a maximally entangled state.
 A key feature in Li et al. \cite{LDM02} is the linearity assumption, that is, both the cost function and the inverse demand function are linear.
 It is well known that linear Cournot games have exactly one equilibrium \cite{Tirole}.
 On the other hand, in nonlinear settings, there may be multiple equilibria, and hence we may not predict the market price.
 A natural question is whether the uniqueness of equilibria is guaranteed in the quantum Cournot duopoly?
 We are trying to answer this question in this paper.
 To quantize the model, we apply Li et al.'s \cite{LDM02} ``minimal'' quantization rules to Cournot duopoly in a nonlinear setting, where there are one symmetric equilibrium and two asymmetric equilibria in the zero entanglement case. \footnote{Several quantum extensions of oligopolistic competition, applying Li et al.'s \cite{LDM02} ``minimal'' quantization rules, have been considered, e.g., the quantum Cournot duopoly game \cite{DLJ03, DJL05}, the quantum Bertrand duopoly game \cite{LK04, QCSD05}, the quantum Stackelberg duopoly game \cite{LK03a, LK05}, and the quantum oligopoly game \cite{LK03b}.}
 We observe the transition of the game from purely classical to fully quantum, as the game's entanglement increases from zero to maximum.
 We show that if the entanglement of the game is sufficiently large, then all asymmetric equilibria vanish and there remains one symmetric equilibrium.
 Furthermore, similar to Li et al. \cite{LDM02}, in the maximally entangled game, the unique symmetric equilibrium is exactly Pareto optimal.
 In other words, the multiplicity of equilibria as well as the dilemma--like situation in the classical Cournot duopoly is completely resolved in our quantum extension.

\section{Classical Cournot Duopoly}
 We consider Cournot competition between two firms, firm $1$ and firm $2$.
 They simultaneously decide the quantity $q_1$ and $q_2$, respectively, of a homogenous product they want to put on the market. Let $P(Q)$ be the inverse demand function, where $Q=q_1+q_2$.
 Each firm $j\in \{1,2\}$ has the common cost function $c(q_j)$.
 Then the firm $j$'s profit can be written as
 \begin{equation}\label{ClaPay}
 u_j(q_1,q_2)=P(Q)q_j-C(q_j).
 \end{equation}
 We assume that 
 \begin{eqnarray*}
 P(Q) &=& a+b-Q, \\
 C(q_j) &=& \frac{1}{4}(q_j-a)^4-q_j^2+bq_j-d,
 \end{eqnarray*}
 where $a,b,d>0$.
 
 Given any $q_2$, we have $\partial^2u_1(q_1,q_2)/\partial q_1^2\le0$. Thus, to maximize her profit, firm $1$ chooses $q_1$ such that $\partial u_1/\partial q_1=0$, that is,
 \begin{equation}\label{FOC1}
 -q_2-(q_1-a)^3+a=0.
 \end{equation}
 
 Similarly, given any $q_1$, firm $2$ chooses $q_2$ such that
 \begin{equation}\label{FOC2}
 -q_1-(q_2-a)^3+a=0.
 \end{equation}
 A pair $(q_1,q_2)$ is a Nash equilibrium iff it solves Eq. (\ref{FOC1}) and (\ref{FOC2}). Then, there are three equilibria, $(q_1,q_2)=(a,a),(a-1,a+1),$ and $(a+1,a-1)$.
 At these equilibria the profits are
 \[
 (u_1(q_1,q_2),u_2(q_1,q_2))=
 \begin{cases}
 (d,d) & \text{if $(q_1,q_2)=(a,a)$} \\
 (3/4-a+d,3/4+a+d) & \text{if $(q_1,q_2)=(a-1,a+1)$} \\
 (3/4+a+d,3/4-a+d) & \text{if $(q_1,q_2)=(a+1,a-1)$}.
 \end{cases}
 \]
  
 However, these equilibria fail to be Pareto optimal.
 The reason why they fail is that both firms can be better off by jointly decreasing their outputs, since $\partial u_i/\partial q_i=0$ and $\partial u_i/\partial q_j=-q_i<0$ at equilibria.
 On the other hand, if the two firms can cooperate and restrict their quantities to
 \[
   q^*\equiv q_1^*=q_2^*=a+2\alpha\beta^{-1}-\frac{1}{2}\alpha^2\beta,
 \]
 where $\alpha\equiv(2/3)^{1/3}$ and $\beta\equiv\left(9a+\sqrt{96+81a^2}\right)^{1/3}$, then they can maximize their joint profit $u_1+u_2$ (obviously $(q^*,q^*)$ is Pareto optimal).
 For example,
 \[
 u_1(q^*,q^*)=u_2(q^*,q^*)= 7/4+d,
 \]
 for $a=3$. Thus, the joint profit at $(q^*,q^*)$ is greater than that of any equilibrium.
 
 With regard to the asymmetric equilibria, the situation is similar to that of chicken game if we correspond the equilibria $(a-1,a+1),(a+1,a-1)$ respectively to the equilibria $(\mathit{cooperate,defect}),$ $(\mathit{defect,cooperate})$ in chicken game.
 Below we will see that, as the measure of entanglement goes to infinitely large in a quantum form of Cournot competition, the unique equilibrium comes to be optimal, as if the unique cooperative equilibrium is attained in chicken game \cite{FH07}.

\section{Quantum Cournot Duopoly}
 To model Cournot duopoly on a quantum domain, we follow Li et al.'s ``minimal'' extension, which utilizes two single--mode electromagnetic fields, of which the quadrature amplitudes have a continuous set of eigenstates.
 The tensor product of two single--mode vacuum states $\ket{vac}_1\otimes \ket{vac}_2$ is identified as the starting state of the Cournot game,
 and the state consequently undergoes a unitary entanglement operation $\hat{J}(\gamma)\equiv \exp\{-\gamma(\hat{a}_1^\dagger\hat{a}_2^\dagger-\hat{a}_1\hat{a}_2)\}$, in which $\hat{a}_1$ and $\hat{a}_2$ ($\hat{a}_1^\dagger$ and $\hat{a}_2^\dagger$) are the annihilation (creation) operators of the electromagnetic field modes.
 The operation is assumed to be known to both firms and to be symmetric with respect to the interchange of the two field modes.
 The resultant state is given by $\ket{\psi_i}\equiv \hat{J}(\gamma)\ket{vac}_1\otimes\ket{vac}_2$.
 Then firm $1$ and firm $2$ execute their strategic moves via the unitary operations $\hat{D}_1(x_1)\equiv \exp\{x_1(\hat{a}_1^\dagger-\hat{a}_1)/\sqrt{2}\}$ and $\hat{D}_2(x_2)\equiv \exp\{x_2(\hat{a}_2^\dagger-\hat{a}_2)/\sqrt{2}\}$, respectively, which correspond to the quantum version of the strategies of the Cournot game.
 The final measurement is made, after these moves are finished and a disentanglement operation $\hat{J}(\gamma)^\dagger$ is carried out.
 The final state prior to the measurement, thus, is $\ket{\psi_f}\equiv \hat{J}(\gamma)^\dagger\hat{D}_1(x_1)\hat{D}_2(x_2)\hat{J}(\gamma)\ket{vac}_1\otimes\ket{vac}_2$.
 The measured observables are $\hat{X}_1\equiv(\hat{a}_1^\dagger+\hat{a}_1)/\sqrt{2}$ and $\hat{X}_2\equiv(\hat{a}_2^\dagger+\hat{a}_2)/\sqrt{2}$, and the measurement is done by the homodyne measurement with an infinitely squeezed reference light beam (i.e., the noise is reduced to zero). 
 When quantum entanglement is not present, namely $\gamma=0$, this quantum structure faithfully represent the classical game, and the final measurement provides the original classical results: $q_1\equiv\braket{\psi_f|\hat{X}_1|\psi_f}=x_1$ and $q_2\equiv \braket{\psi_f|\hat{X}_2|\psi_f}=x_2$.
 Otherwise, namely when quantum entanglement is present, the quantities the two firms will produce are determined by
 \begin{eqnarray*}
 q_1 &=& x_1\cosh \gamma +x_2\sinh \gamma,\\
 q_2 &=& x_2\cosh \gamma +x_1\sinh \gamma.
 \end{eqnarray*}
 Note that the classical model can be recovered by choosing $\gamma$ to be zero, since the two firms can directly decide their quantities. On the other hand, both $q_1$ and $q_2$ are determined by $x_1$ and $x_2$ when $\gamma\neq 0$. It leads to the correlation between the firms.
 
 Substituting $q_j$ into Eq. (\ref{ClaPay}) provides the quantum profits $u^Q_j$ for firm $j$:
 \[
 u_j^Q(x_1,x_2) = P(x_1,x_2)(x_j \cosh \gamma + x_i \sinh \gamma)-C(x_j,x_i).
 \]
 where $i\ne j$ and
 \begin{eqnarray*}
  P(x_1,x_2) & = & a + b - e^\gamma (x_1+x_2),\\
  C(x_j,x_i) & = &  \frac{1}{4}(x_j\cosh\gamma +x_i\sinh\gamma -a)^4 \\
  && \qquad-(x_j\cosh\gamma + x_i\sinh \gamma)^2 + b(x_j\cosh \gamma +x_i\sinh \gamma) -d.
 \end{eqnarray*}
 
 Similar to the classical game, we also have $\partial^2u_j(x_j,x_i)/\partial x_j^2\le 0$ for any $x_i$. To maximize her profit, thus, firm $j$ chooses $x_j$ such that $\partial u_j/\partial x_j=0$, that is,
 \begin{equation}\label{FOC3}
   - x_j\sinh{2\gamma}
   - x_i\cosh{2\gamma}
   - (x_j \cosh{\gamma} + x_i \sinh{\gamma} -a)^3\cosh{\gamma} +a \cosh{\gamma}=0.
 \end{equation}
 Solving Eq. (\ref{FOC3}) for both firms provides the quantum Nash equilibria, and the symmetric one is uniquely given as
 \begin{equation}\label{symNash}
 x^*(\gamma)\equiv x^*_1=x^*_2 = \frac{a}{ e^{\gamma}}
 + \text{sech}{\gamma}\cdot\alpha\,\eta^{-1}
 -\frac{1}{2}\alpha^2
 \frac{\eta}{e^{\gamma}},
 \end{equation}
 where $\eta \equiv 
 \left(9 a \tanh \gamma + \sqrt{12 e^{3 \gamma} \text{sech}^3{\gamma}+81 a^2\tanh^2{\gamma}}\right)^{1/3}$. 
 As easily seen from Eq. (\ref{symNash}), the quantity produced by each firm in the equilibrium, equals to $e^\gamma x^*(\gamma)$,  monotonically increases and approaches to the Pareto optimal one $q^*$, as the entanglement $\gamma$ increases.
 In fact, given $\lim_{\gamma \rightarrow \infty}{\tanh{\gamma}}=1$ and $\lim_{\gamma \rightarrow \infty}{e^\gamma\text{sech}\gamma}=2$, we have $\lim_{\gamma \rightarrow \infty} \eta  =\beta$, and thus, 
 \[
 \lim_{\gamma \rightarrow \infty}{e^\gamma x^*(\gamma)} = q^*.
 \]
 
\begin{figure}[ht]
   \begin{center}
     \begin{tabular}{c}
       \includegraphics[width=4.5cm]{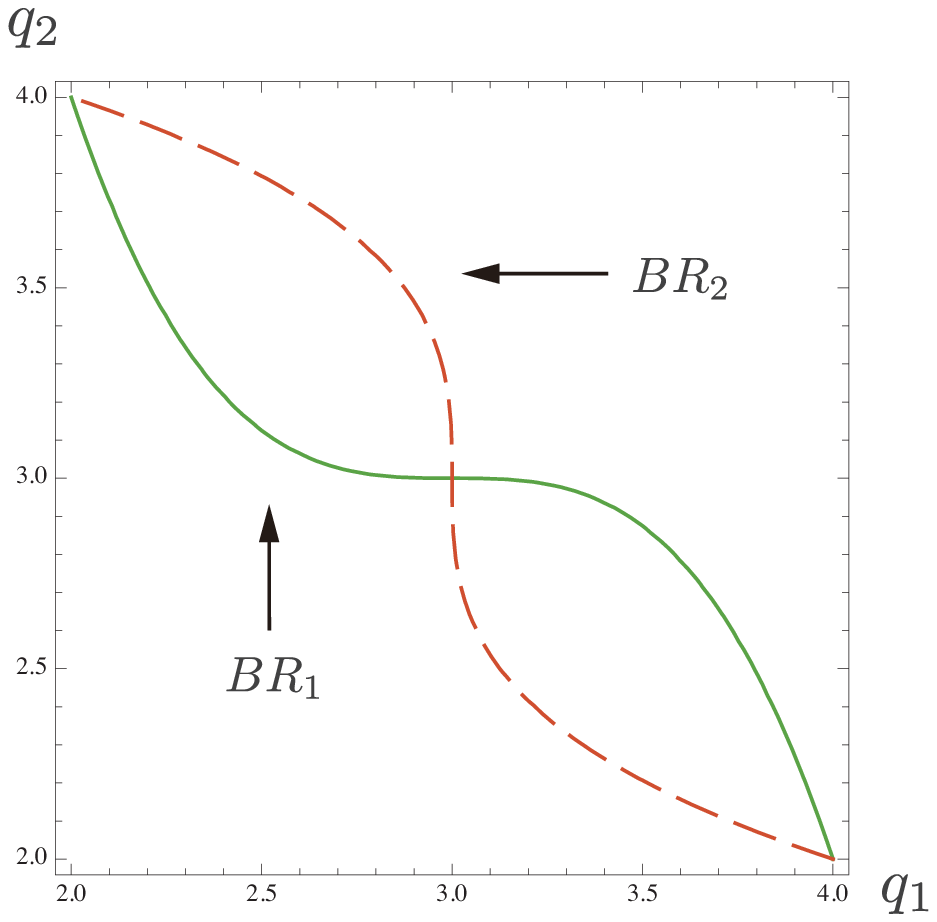}\\
       (i) $\gamma=0$
     \end{tabular}
     \quad
     \begin{tabular}{c}
       \includegraphics[width=4.5cm]{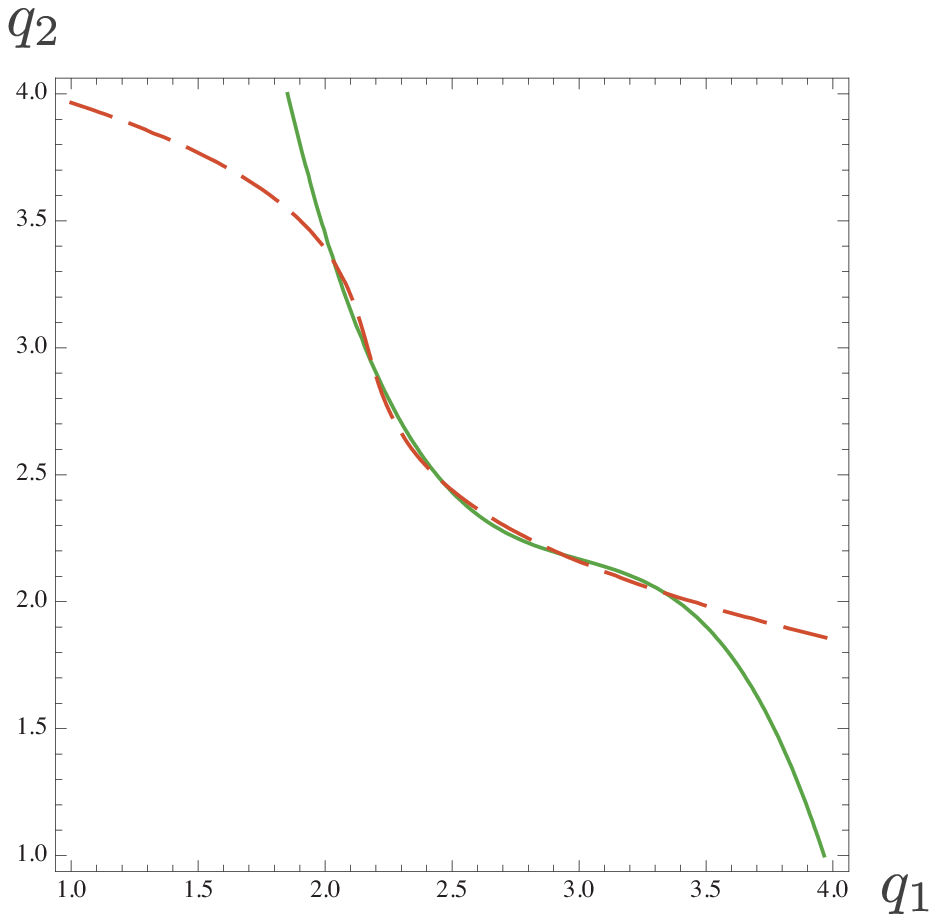}\\
       (ii) $\gamma=.285$
     \end{tabular}
     \quad
     \begin{tabular}{c}
       \includegraphics[width=4.5cm]{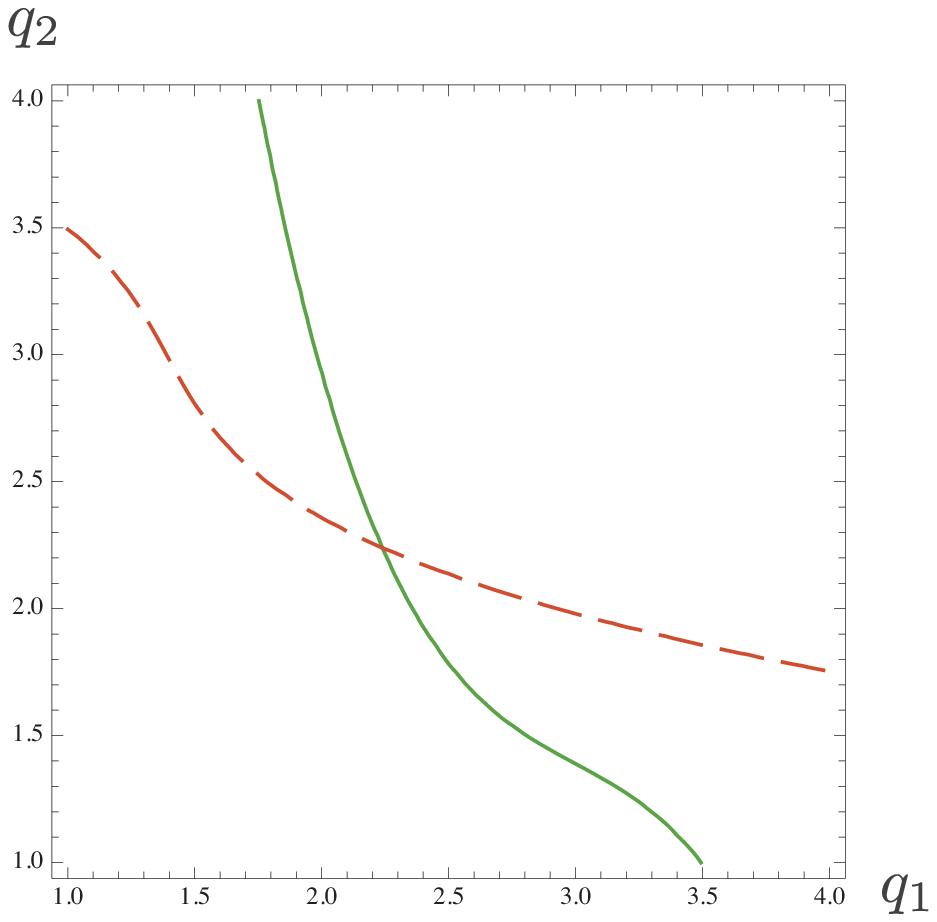}\\
       (iii) $\gamma=.6$
     \end{tabular}
     \caption{\label{figBR}
     $BR_1$ and $BR_2$ for $a=3$.
    }
   \end{center}
 \end{figure}

 As we have observed above, in addition to the symmetric one there are two asymmetric equilibria in the classical model.
 Hence, it is expected that the quantum model also possesses asymmetric equilibria at least as far as the entanglement is not too large.
 In fact, we can see that this conjecture is valid for the case of $a=3$ as follows.
 By substituting $x_j$ with $q_j$, Eq. (\ref{FOC3}) can be rewritten as
 \begin{equation}\label{BR}
   BR_j: \quad (a + q_j - q_i -(q_j -a)^3)\cosh{\gamma} -e^{\gamma} q_j=0.
 \end{equation}
 $BR_j$ is a locus of quantities, which is determined by firm $j$'s best response strategy $x_j$ to the opponent's strategy $x_i$. Each intersection of $BR_1$ and $BR_2$ represents quantities produced in some equilibrium.
 Fig. \ref{figBR} depicts $BR_1$ and $BR_2$ for $\gamma=0, .285,\text{ and } .6$, respectively.
 \noindent 
 Fig. \ref{figBR} displays that the number of equilibria varies as $\gamma$ changes.
 Fig. \ref{figBR}(i) corresponds to the classical model, where there are three equilibria.
 Fig. \ref{figBR}(ii), in which five equilibria exist, shows the possibility that the number of equilibria increases by the existence of the entanglement.
 In Fig. \ref{figBR}(iii), asymmetric equilibria disappear and the only one symmetric equilibrium remains. 
 Evidently from these figures, there is the possibility of multiple equilibria even if the entanglement exists.
 However, we can prove that asymmetric equilibria vanish when $\gamma$ goes large.

\begin{pro}
For a sufficiently large $\gamma$, $(x^*(\gamma),x^*(\gamma))$ is the unique equilibrium. 
\end{pro}

 It is worth pointing out that the quantifier ``sufficiently large'' in the proposition is not so restrictive by the following reason.
 To obtain the proposition, we use the fact that $\mathrm{sech}\gamma/e^\gamma\to 0$ as $\gamma \to \infty$.
 Since $\mathrm{sech}\gamma/e^\gamma$ converges very quickly, the lower bound for $\gamma$, which guarantees the uniqueness of equilibria, is not so large.
 For instance, any asymmetric equilibrium cannot exist for $\gamma > \gamma_2 \simeq .296$ when $a = 3$ (as we will see below in Fig. \ref{figPayoff}). 

 Finally, we consider the transition of equilibria of the game from purely classical to fully quantum, as $\gamma$ increases from zero to infinity.
 Fig. \ref{figPayoff} depicts the transition process for the case of $a=3$ and $d=10$: the number of equilibria changes, as $\gamma$ grows large, from 3 to 5, from 5 to 3, and from 3 to 1 at last.
 More precisely, there are two thresholds, namely $\gamma_1\simeq .255$ and $\gamma_2\simeq .296$: for $0\leq \gamma<\gamma_1$, there are three equilibria; for $\gamma_1\le \gamma <\gamma_2$, there are five; for $\gamma=\gamma_2$, there are three; and for $\gamma_2<\gamma$, the symmetric (and unique) one remains.
 The horizontal line at 11.75 ($= 7/4 +d$) represents the half of the maximum joint profit, and the line at 10 represents the profit at the symmetric Nash equilibrium of the classical Cournot game.
 As easily seen from the figure, asymmetric equilibria vanish and the unique symmetric equilibrium monotonically approaches to the optimal one as $\gamma$ goes large.
 
 \begin{figure}[ht]
   \begin{center}
     \includegraphics[width=8cm]{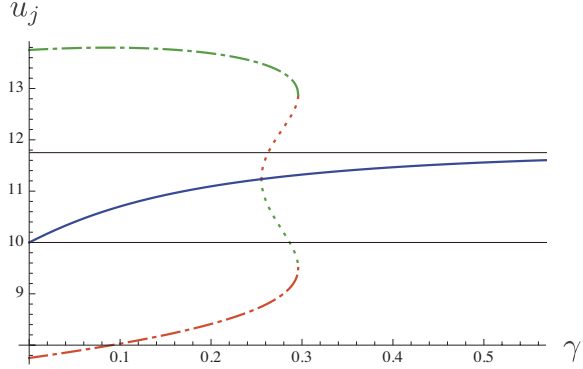}
     \caption{\label{figPayoff} The profits at quantum Nash equilibria as a function of the entanglement parameter $\gamma$, where $a=3$ and $d=10$.
     }
   \end{center}
 \end{figure}

\section*{Appendix: Proof of the Proposition}
 Given a sufficiently large $\gamma$, let $(q_1,q_2)$ be equilibrium outputs. Suppose to the contrary, $q_2-q_1=\delta\neq 0$. Then, Eq. (\ref{BR}) implies:
 \begin{eqnarray}
 	& &(a-\delta-A^3)\cosh \gamma-e^\gamma q_1=0, \label{FOC4} \\
 	& &(a+\delta -(A+\delta)^3)\cosh \gamma -e^\gamma (q_1+\delta) =0, \label{FOC5} 
 \end{eqnarray}
 where $A \equiv q_1-a$.
 Subtracting Eq. (\ref{FOC4}) from Eq. (\ref{FOC5}), we have
 \[
	(2\delta -B)\cosh \gamma -\delta e^\gamma= 0,
 \]
 where $B \equiv 3A^2\delta+3A\delta^2+\delta^3$.
 It implies
 \[
 	B\cosh \gamma = \delta(2\cosh \gamma-e^\gamma) = \delta e^{-\gamma}.
 \]
 Since we assume that $\delta\neq 0$, $B/\delta = e^{-\gamma}/\cosh \gamma$, that is,
 \begin{equation}
 	\delta^2+3A\delta+3A^2-\frac{\mathrm{sech}\gamma}{e^\gamma}=0. \label{delta}
 \end{equation}
 It is necessary for Eq. (\ref{delta}) having a real solution that 
 \[
 	\frac{4}{3}\frac{\mathrm{sech}\gamma}{e^\gamma}\ge A^2,
 \]
 which implies that $A$ must be sufficiently close to zero since $\mathrm{sech}\gamma/e^\gamma\to 0$ as $\gamma\to \infty$.
 The solution of Eq. (\ref{delta}) is given by
 \[
 	\delta=\frac{-3A\pm\sqrt{-3A^2+4\frac{\mathrm{sech}\gamma}{e^\gamma}}}{2} \simeq 0.
 \]

 On the other hand, Eq. (\ref{FOC4}) implies that 
 \[
 	\delta =a-A^3-q_1e^\gamma\mathrm{sech}\gamma=a-A^3-(A+a)e^\gamma\mathrm{sech}\gamma.
 \]
 Since $e^\gamma\mathrm{sech}\gamma\to 2$ and $A\to 0$ as $\gamma\to \infty$, we obtain
 \[
	\delta\simeq -a,
 \]
 which implies that $\delta$ is bounded away from zero, a contradiction.
 Thus, for a sufficiently large $\gamma$, any equilibrium must be symmetric.
 However, $(x^*(\gamma),x^*(\gamma))$ is the unique symmetric solution of Eq. (\ref{FOC3}).
 $\Box$

\end{document}